\documentclass[conference]{IEEEtran}

\IEEEoverridecommandlockouts
\usepackage{verbatim}
\usepackage{longtable}
\usepackage{adjustbox}
\usepackage{float}
\usepackage{amsmath}
\usepackage{amsfonts}
\usepackage{amssymb}
\usepackage{multirow}
\usepackage[T1]{fontenc}
\usepackage{amsmath,amssymb,amsfonts}
\usepackage{fancyhdr}
\usepackage[ruled,vlined,linesnumbered]{algorithm2e}
\usepackage{graphicx}
\usepackage{textcomp}
\usepackage{xcolor}
\usepackage{subcaption}
\usepackage{algpseudocode}
\usepackage{enumitem}
\usepackage{float}
\usepackage{graphicx}
\usepackage[normalem]{ulem}
\usepackage{url}
\useunder{\uline}{\ul}{}
\begin{document}

\title{Bridging Immutability with Flexibility: A Scheme for Secure and Efficient Smart Contract Upgrades}


\author{\IEEEauthorblockN{Tahrim Hossain$^{1}$, Sakib Hassan$^{2}$, Faisal Haque Bappy$^{3}$, Muhammad Nur Yanhaona$^{4}$,\\Tarannum Shaila Zaman$^{5}$, and Tariqul Islam$^{6}$}
\IEEEauthorblockA{
$^{1, 3, 6}$ Syracuse University, USA;
$ ^{2}$ University Of Dhaka, Bangladesh;\\
$ ^{4}$ BRAC University, Bangladesh;
$ ^{5}$ University of Maryland, Baltimore County\\
Email: mhossa22@syr.edu, skbhssn.sh@gmail.com, fbappy@syr.edu,  nur.yanhaona@bracu.ac.bd,\\zamant@umbc.edu, and mtislam@syr.edu} 
}

\maketitle

\thispagestyle{fancy}
\lhead{This work has been accepted at the IEEE International Conference on Blockchain and Cryptocurrency (ICBC 2025)}
\pagestyle{plain}

\begin{abstract}
The emergence of blockchain technology has revolutionized contract execution through the introduction of smart contracts. Ethereum, the leading blockchain platform, leverages smart contracts to power decentralized applications (DApps), enabling transparent and self-executing systems across various domains. While the immutability of smart contracts enhances security and trust, it also poses significant challenges for updates, defect resolution, and adaptation to changing requirements. Existing upgrade mechanisms are complex, resource-intensive, and costly in terms of gas consumption, often compromising security and limiting practical adoption. To address these challenges, we propose \textit{FlexiContracts+}, a novel scheme that reimagines smart contracts by enabling secure, in-place upgrades on Ethereum while preserving historical data without relying on multiple contracts or extensive pre-deployment planning. \textit{FlexiContracts+} enhances security, simplifies development, reduces engineering overhead, and supports adaptable, expandable smart contracts. Comprehensive testing demonstrates that \textit{FlexiContracts+} achieves a practical balance between immutability and flexibility, advancing the capabilities of smart contract systems.
\end{abstract}

\section{Introduction}
Smart contracts are self-executing programs that run on the blockchain platform and automatically enforce predefined rules and conditions \cite{buterin2013ethereum}. This concept, originally proposed by Nick Szabo \cite{szabo1996smart}, enables developers to create decentralized applications (DApps) without requiring trusted intermediaries. Ethereum smart contracts are designed to be immutable once deployed to the blockchain to ensure security by providing reliable execution logic \cite{qasse2024immutable} and enabling audit trails for transparent and verifiable interactions \cite{tsankov2018securify}. However, immutability directly conflicts with the reality of software development - bugs need fixing, security vulnerabilities require patching, and business requirements evolve \cite{hofmann2017immutability}. Two major incidents, the 2016 DAO hack \cite{daoxploit2016} and the 2017 Parity wallet bug \cite{paritywalletmultisighack2017} highlight the risks posed by smart contract immutability.
This tension between immutability and upgradeability affects smart contract implementations across financial services, healthcare systems, and supply chain management \cite{tapscott2016Blockchain, lim2021literature, kuo2017blockchain}.

Traditional methods for smart contract upgradability fall into two categories: redeployment with data migration and design patterns, each introducing additional complexity and potential security risks. Redeployment with data migration requires deploying a new contract and transferring state, a labor-intensive process that risks data loss, address changes, and user disruption. While automated tools assist with state extraction and transfer \cite{ayub2023storage, ayub2024sound}, this method still lacks support for seamless in-place upgrades.

Design patterns like Proxy \cite{cyfrin_proxy}, Diamond \cite{diamondPattern}, Eternal Storage \cite{openzeppelin_eternal_storage}, and Metamorphic contracts \cite{mixbytes_metamorphic_contracts}, on the other hand,  enable modular upgrades but require intricate instrumentation, advanced storage management, and deep knowledge of Ethereum’s execution environment, making them error-prone and susceptible to vulnerabilities. Several studies have focused on developing taxonomies of these security issues \cite{bodell2023proxy,li2024characterizing} and frameworks for detecting them \cite{ruaronot}, but they largely fall short of providing proactive methods to mitigate these vulnerabilities, leaving critical gaps in ensuring secure and efficient contract upgrades. Both approaches also pose centralization risks, often relying on a single entity for upgrades, undermining decentralization and creating a single point of failure vulnerable to misuse or error.

In response to these challenges, we present \textit{FlexiContracts+}, a novel scheme that reconceptualizes smart contracts by realizing upgradeability as a native, protocol-level capability that eliminates the need for redeployment and complex instrumentation. \textit{FlexiContracts+} embeds on-chain governance and automated storage reorganization into the contract architecture, resulting in a single-contract model that enables transparent stakeholder-driven upgrades while aligning contract state with updated logic to ensure state integrity. This single-contract model avoids delegated logic execution, common in proxy-based patterns simplifying contract development and enhancing security. Through comprehensive analysis and evaluation, we demonstrate the effectiveness of \textit{FlexiContracts+} in overcoming the limitations of traditional upgrade mechanisms and setting a new standard for smart contract upgradability.

The core contributions of this paper are as follows.
\begin{itemize}
    \item We introduce an upgrade framework that reconceptualizes smart contracts by embedding on-chain governance and automated storage reorganization into the contract architecture, enabling seamless in-place upgrades as a native capability.

    \item We implement proof-of-concept of \textit{FlexiContracts+} by extending Ethereum's architecture with a governance-aware account structure that enables stakeholder voting, an automated storage reorganization layer that preserves state integrity during upgrades, and custom transaction types that orchestrate the entire upgrade workflow.
    
    \item Our proof-of-concept showed competitive performance against four existing schemes. Also, gas cost analysis across six diverse smart contract configurations shows \textit{FlexiContracts+} requires a negligible amount of extra gas, validating its efficient upgrade logic.
    
    
\end{itemize}

We structure the remainder of this paper as follows: in
Section 2, we critically review related work. In Section 3, we present the system architecture, detailing \textit{FlexiContracts'+} core components. In Section 4, we assess \textit{FlexiContracts+} via performance metrics, and gas cost analysis. We conclude the paper in Section 5 summarizing the key findings.

\section{Related Work}

Recent research has focused on balancing flexibility, security, and decentralization in smart contract upgradability, while still facing significant trade-offs. Initial studies have explored off-chain mechanisms for smart contract upgradability \cite{fourTier}, but these approaches compromise decentralization, reduce transparency, and introduce security risks, making them incompatible with blockchain’s trustless nature. EVMPatch, proposed by Rodler et al., enables upgrades through bytecode-level patching by automatically converting contracts into the delegatecall-proxy pattern \cite{evmpatch}. While this removes the need for manual rewrites, it complicates contract verification and auditing, raising concerns about transparency and potential vulnerabilities.

Traditional upgrade methods often involve deploying new contracts and manually migrating state, a process fraught with challenges such as data loss, errors, and high costs, especially when handling complex data structures. While Ayub et al. have introduced automated tools to analyze and extract contract states to ease migration \cite{ayub2023storage,ayub2024sound}, they focus on external mechanisms and post-deployment extraction, offering limited support for seamless in-place upgrades.

The Ethereum community has developed several design patterns, including Proxy \cite{cyfrin_proxy}, Diamond \cite{diamondPattern}, Eternal Storage \cite{openzeppelin_eternal_storage}, and Metamorphic contracts \cite{mixbytes_metamorphic_contracts}, to address upgradability. While these patterns enable flexibility and state preservation, they require complex source code instrumentation and multi-contract systems, leading to high gas costs and security vulnerabilities. Further studies have analyzed these patterns, with Bodell et al. providing a taxonomy of upgradable smart contracts (USCs) and detecting vulnerabilities through source code analysis \cite{bodell2023proxy}. Building on this, Li et al. introduced USCDetector, which identifies vulnerabilities directly from bytecode, making it suitable for auditing deployed contracts without requiring source code access \cite{li2024characterizing}. Ruaro et al. further highlight storage collisions in proxy-based systems using the CRUSH framework, exposing risks such as privilege escalation and frozen funds \cite{ruaronot}.Despite these advancements, there remains a pressing need for a secure and efficient upgrade framework that eliminates the need for additional complex instrumentation while mitigating vulnerabilities and integrating governance mechanisms to enhance transparency and usability in smart contract systems.
\section{System Architecture}
\begin{figure*} [tbh]
    \centering
    \includegraphics[width=0.75\textwidth]{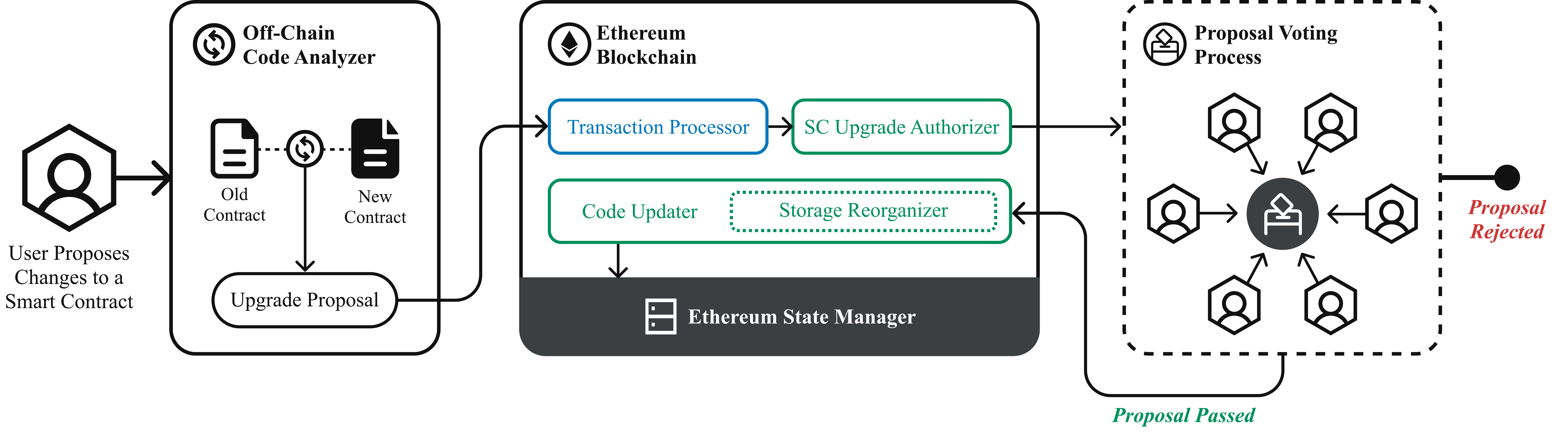}
    \caption{System Workflow} 
    \label{fig:workflow} 
\end{figure*}

This section outlines the interconnected components of \textit{FlexiContracts+} and their integration within the workflow shown in Fig.~\ref{fig:workflow}, highlighting how they coordinate governance and dynamic storage reorganization to streamline upgrades.

\subsection{Off-Chain Code Analyzer}
We design the Off-Chain Code Analyzer to initiate the upgrade process by comparing existing and proposed smart contract source code versions, analyzing changes in storage layout, and evaluating their impact on data organization. Ethereum’s storage consists of $2^{256}$ 32-byte slots managed by the Storage Trie, where slots are allocated based on variable declaration order and size. Complex structures like arrays and mappings use hash-based indexing, making upgrades susceptible to data misalignment and corruption. To address this, the analyzer generates precise reorganization instructions to align the existing storage with the updated layout. These instructions, along with governance parameters, are encapsulated into our custom transaction types and embedded in our modified state account structure. These transactions are subsequently forwarded to the Transaction Processor, which processes them based on their type.

\subsection{Transaction Processor}
We manage governance and upgrade-related workflows through our Transaction Processor, which interprets and handles our custom transaction types. When a transaction initiates a deployment, we initialize the contract along with its associated governance configuration. For transactions directed at existing contracts, we differentiate between proposals for contract upgrades, interactions with the current contract logic, votes cast by stakeholders, and the final application of approved changes. We distinguish between these intents using newly added parameters in our custom transaction types, along with metadata from the modified state account structure. We strategically separate the initiation of smart contract upgrades from their implementation. This division into distinct transactions helps reduce uncertainty around resource use, enables the scheduling of upgrades under favorable network conditions to minimize gas costs, and alleviates the financial burden on proposers by spreading costs across different phases. The Transaction Processor forwards the transactions to the SC Upgrade Authorizer for validation and execution.

\subsection{SC Upgrade Authorizer}
We have designed the SC Upgrade Authorizer in \textit{FlexiContracts+} to streamline the governance workflow for smart contract upgrades. This component utilizes the modified account structure, particularly the Proposal Trie and Ballot Trie which are both applications of Merkle Patricia Tries \cite{jezek2021ethereum} used for organizing data in a hierarchical, tamper-proof manner. These structures allow us to securely manage, track, and validate proposals and votes, enhancing the transparency and consistency of the upgrade process.

When a \textit{Proposal} transaction is submitted, we validate the stakeholder’s authorization and ensure no other active proposal exists to maintain governance consistency. Upon successful validation, the Proposal Trie is updated with the new proposal and relevant metadata. The timeout mechanism, integrated into the modified account structure, ensures that unresolved proposals expire after a predefined period, preventing indefinite blocking of the governance process. Once registered, the voting phase is activated, allowing stakeholders to review the proposal and cast their votes.

After activating the voting phase, stakeholders can submit \textit{Vote} transactions to decide on the proposal. We validate each vote to ensure the voter is authorized and the proposal is active. Valid votes are recorded in the Ballot Trie, a secure data structure that links each vote to the corresponding proposal and stakeholder. We maintain a running tally of the votes to monitor overall support or opposition. Once the votes meet the required threshold, the proposal is marked as approved in the Proposal Trie and moves to the application phase.

During the upgrade workflow, stakeholders can trigger the execution of a smart contract by submitting an \textit{Execute} transaction. Before proceeding, we check whether the voting threshold to deactivate execution has been reached. This threshold is a crucial governance parameter that allows stakeholders to temporarily halt the execution of a potentially faulty contract version while deliberation on an upgrade proposal is ongoing. If the threshold is not met, we proceed with the contract's execution, ensuring normal operations continue while respecting governance decisions.

Once a proposal is approved, stakeholders can submit an \textit{Apply} transaction to implement the changes. We validate the request to confirm approval and trigger the storage reorganization process to ensure compatibility with the upgraded contract. Using the Code Updater, we deploy the new smart contract code to the blockchain and finalize the proposal state.

\subsection{On-Chain Storage Reorganizer}
The On-Chain Storage Reorganizer is a critical component in \textit{FlexiContracts+}, ensuring seamless upgrades by aligning the storage layout with updated smart contract code. The reorganizer processes reorganization instructions during the application phase by iterating through each instruction and determining the appropriate reorganization strategy based on the data type and encoding. Below, we describe the reorganization strategies for value types, dynamic arrays, and bytes and strings.

\subsubsection{Reorganization of value types and fixed size arrays}
Value types like uint256, boolean, and fixed-size arrays are stored sequentially in contiguous slots, optimizing for size and alignment. We have implemented inplace reorganization for value types and fixed-size arrays. Our reorganization process involves copying data from its previous slot to the new one, with appropriate offset adjustments.
\subsubsection{Reorganization of dynamically-sized arrays}
Dynamic arrays store their length in a primary slot and their elements in slots derived from the keccak256 hash of the primary slot's index. This storage pattern accommodates variable-length data but introduces complexity during upgrades. We handle this by first copying the array's size to a new primary slot, followed by relocating the array's elements to slots recalculated based on the updated layout. 
\subsubsection{Reorganization of bytes and strings}
Bytes and strings follow a unique encoding pattern. Short data (under 32 bytes) is stored entirely in a single slot, while longer data uses the primary slot for length and hashes the data to separate slots. In the reorganization algorithm we directly copy the primary slot. We then determine whether the data is short or long. For long data, we recalculate the new positions of the elements and update them accordingly.

\section{Evaluation} 
This section evaluates the effectiveness of \textit{FlexiContracts+} in addressing smart contract upgradability challenges through a comprehensive analysis. We developed a proof-of-concept\footnote{Available at \url{https://github.com/FlexiContract/Geth-FlexiContract}} by modifying the Go implementation of Ethereum (Geth\cite{geth}), focusing on code complexity and gas cost comparison to highlight its advantages over existing schemes.

\subsection{Code Complexity Analysis}
We analyzed the performance of \textit{FlexiContracts+} against traditional patterns like Proxy, Eternal Storage, Diamond, and Metamorphic contracts, focusing on complexity.
\begin{figure} [tbh]
  \centering
  \begin{subfigure}[b]{0.48\columnwidth}
    \centering
    \includegraphics[width=\textwidth, height=1.5in]{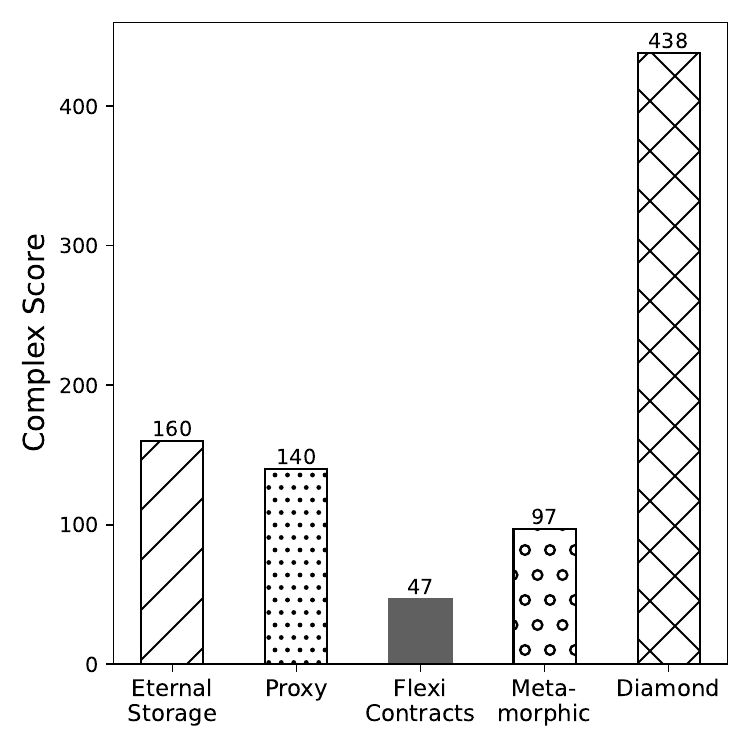}
    \caption{Complex score of ERC-20}
    \label{fig:design_pattern_comparison_ERC_20}
  \end{subfigure}
  \hfill
  \begin{subfigure}[b]{0.48\columnwidth}
    \centering
    \includegraphics[width=\textwidth, height=1.5in]{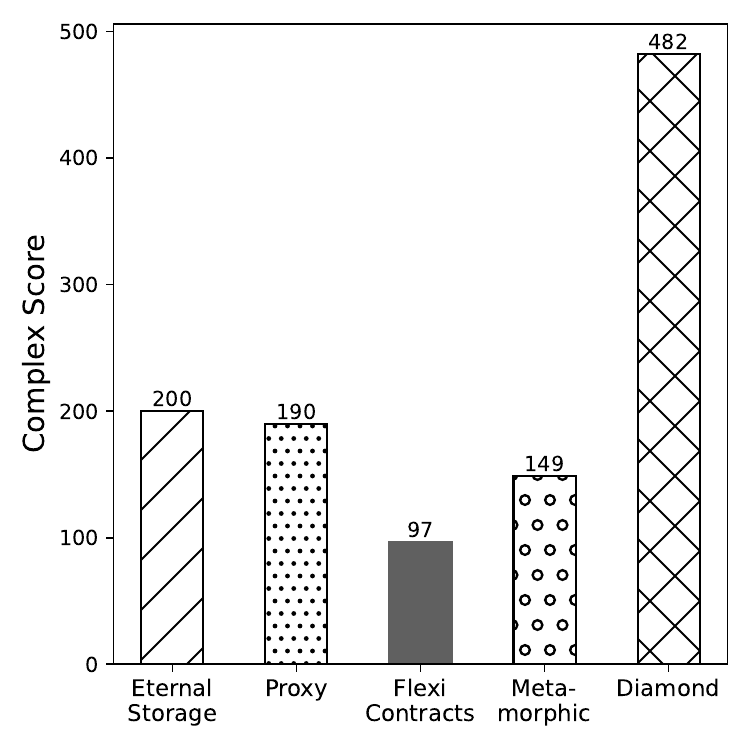}
    \caption{Complex score of ERC-721}
    \label{fig:design_pattern_comparison_ERC_721}
  \end{subfigure}
  
  \vspace{1em}
  
  \begin{subfigure}[b]{0.48\columnwidth}
    \centering
    \includegraphics[width=\textwidth,height=1.5in]{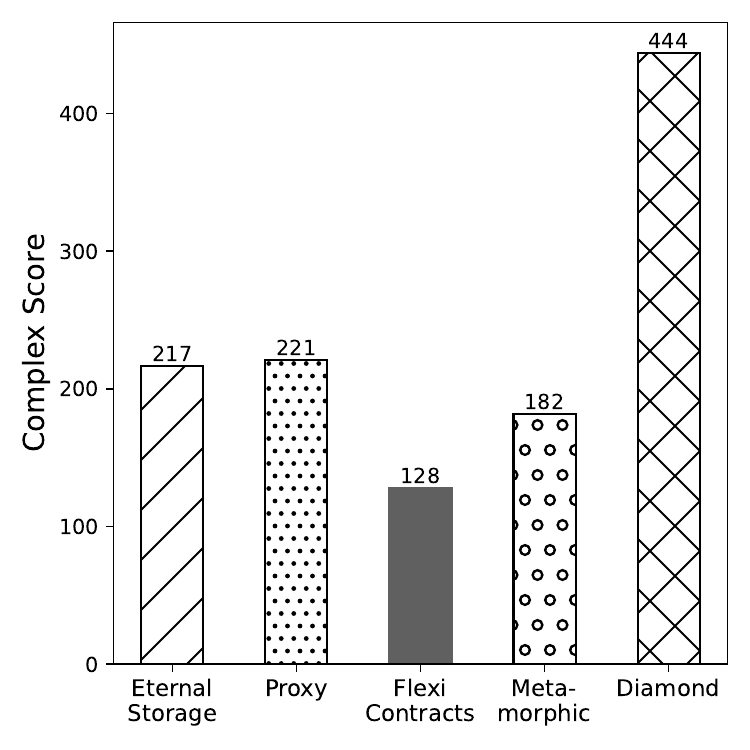}
    \caption{Complex score of ERC-1155}
    \label{fig:design_pattern_comparison_ERC_1155}
  \end{subfigure}
  \hfill
  \begin{subfigure}[b]{0.48\columnwidth}
    \centering
    \includegraphics[width=\textwidth,height=1.5in]{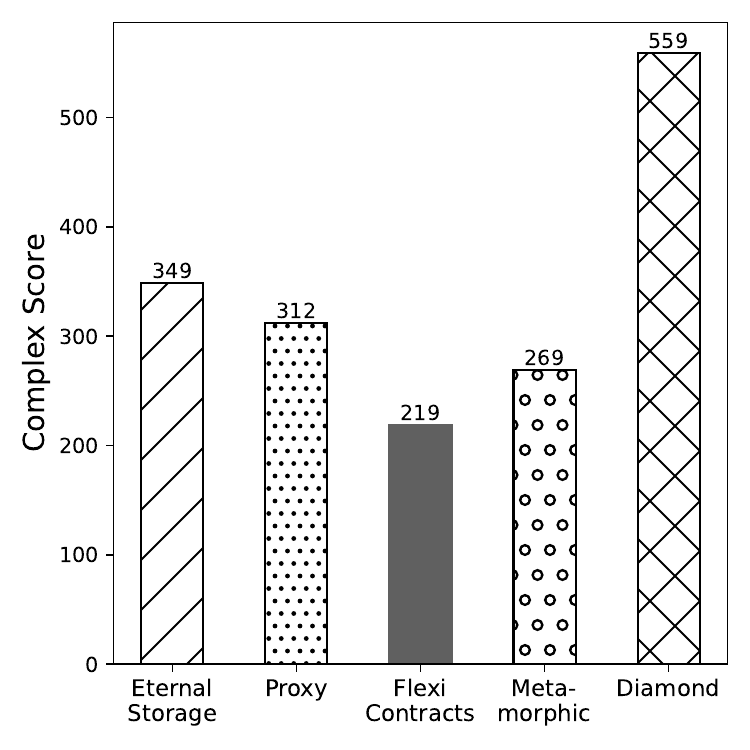}
    \caption{Complex score of ERC-777}
    \label{fig:design_pattern_comparison_ERC_777}
  \end{subfigure}
  
  \caption{Comparison of complex score between Eternal Storage, Proxy, Metamorphic, Diamond Pattern and \textit{FlexiContracts+}}
  \label{fig:overallComp}
\end{figure}
Our analysis included implementations of four popular token standards: ERC-20, ERC-721, ERC-1155, and ERC-777, each featuring the required basic functionalities. We also included a base contract version upgraded seamlessly with \textit{FlexiContracts+}\footnote{Performance comparison data available at \url{https://github.com/FlexiContract/Performance-Analysis-Data}}.

We evaluated complexity using the Solidity metrics tool \cite{soliditymetrics}, comparing different design patterns. As shown in Fig. \ref{fig:overallComp}, the complexity score, based on the number of logic contracts, interfaces, and lines of code, demonstrates that \textit{FlexiContracts+} consistently exhibited significantly lower complexity compared to other patterns. This reduction enhances the developer experience by improving understandability, modularity, and size, simplifying implementation and maintenance. \textit{FlexiContracts+}’s efficient single-contract design and automated reorganization further lower development overhead.

\subsection{Gas Cost Analysis}
\begin{figure}[tbh]
\centering
\begin{minipage}[t]{.46\linewidth}
  \includegraphics[width=\textwidth,height=1.7in]{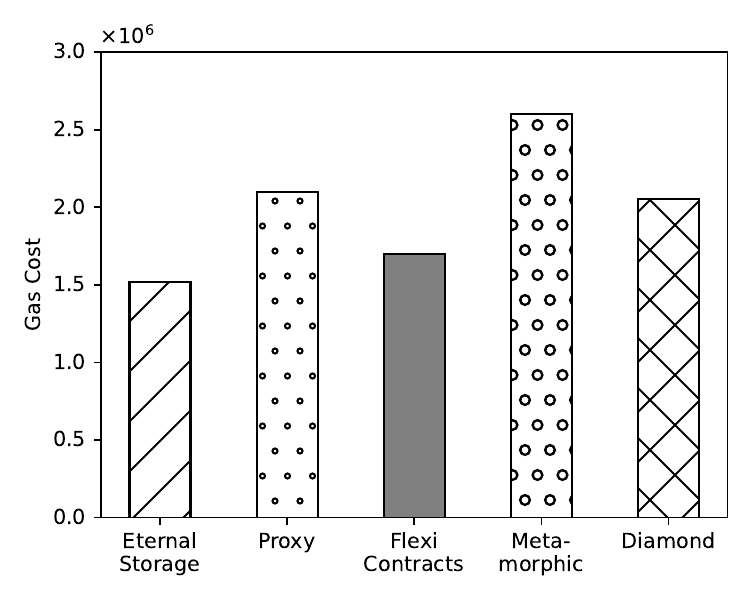}
  \captionof{figure}{Gas cost comparison}
  \label{fig:gas-cost-analysis}
\end{minipage}
\hspace{.03\linewidth}
\begin{minipage}[t]{.46\linewidth}
    
  \includegraphics[width=1.65in,height=1.6in]{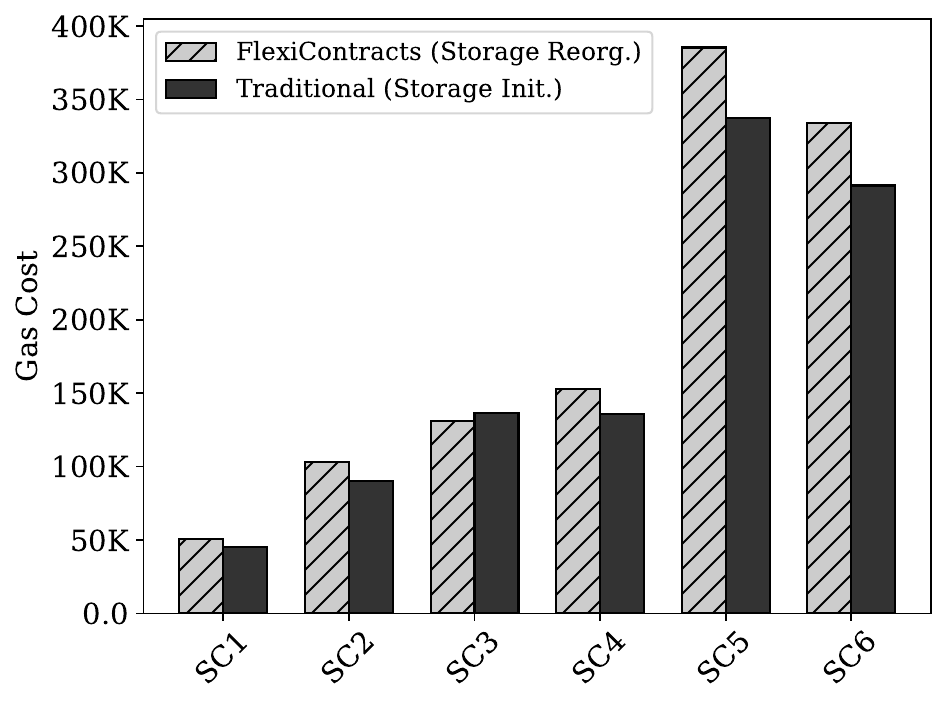}
  \captionof{figure}{Gas costs of reorgs.}
  \label{fig:gas_cost_analysis}
\end{minipage}
\end{figure}

To evaluate the gas costs of \textit{FlexiContracts+}, we compared it with design patterns, as shown in Figure \ref{fig:gas-cost-analysis}. These patterns require deploying a new logic contract and updating the proxy address. However, such comparisons proved less meaningful since these patterns handle storage upgrades differently, requiring new variables to be appended at the end to avoid state corruption, which limits flexibility.
We designed \textit{FlexiContracts+} to support dynamic adjustments to the entire storage layout, offering greater flexibility but making direct comparisons with other methods less meaningful. The lack of standardized governance models for upgrades further complicates gas cost comparisons. Instead, we analyze gas usage by comparing \textit{FlexiContracts+}' storage reorganization with Ethereum’s default method of destroying old contracts, deploying new ones, and migrating data. While traditional approaches are straightforward, they can be costly and risky, requiring careful data handling to avoid loss or corruption. Since \textit{FlexiContracts+}'s On-chain Storage Reorganizer operates outside the EVM, we estimate gas costs by mapping its operations to equivalent EVM opcodes and assigning corresponding gas values.

\begin{table}[tbh]
\centering
\caption{Characteristics of Smart Contracts Used in Gas Cost Analysis}
\label{tab:gas-cost-analysis}
\resizebox{0.8\columnwidth}{!}{%
\begin{tabular}{cccc}
\hline
\textbf{Smart Contract (SC)}       & \textbf{\#Vars} & \textbf{\#Reorgs} & \textbf{Complexity} \\ \hline
\textit{{SC 1}}  & 2                  & 2                    & Simple              \\ \hline
\textit{{SC 2}}  & 2                  & 3                    & Simple              \\ \hline
\textit{{SC 3}}  & 6                  & 10                   & Moderate            \\ \hline
\textit{{SC 4}}  & 5                  & 5                    & Moderate            \\ \hline
\textit{{SC 5}}  & 6                  & 12                   & Complex             \\ \hline
\textit{{SC 6}} & 6                  & 12                   & Complex             \\ \hline
\end{tabular}%
}
\end{table}
To assess gas efficiency, we simulated smart contracts with diverse data types\footnote{Gas cost data available at \url{https://github.com/FlexiContract/Gas_Cost_Data}} and compared the gas costs of storage initialization in new contracts versus reorganization with \textit{FlexiContracts+}, excluding simpler ERC standards for more complex test cases detailed in Table \ref{tab:gas-cost-analysis}. As shown in Figure \ref{fig:gas_cost_analysis}, \textit{FlexiContracts+} incurred a modest gas overhead of 12.58\% to 14.30\%, averaging 10.78\%, while supporting in-place upgrades, preserving addresses and states, reducing data loss risks, and enabling dynamic storage reorganization without append-only constraints.

\section{Conclusion}
In this paper, we introduced \textit{FlexiContracts+}, a scheme that enables secure, efficient, and scalable in-place upgrades for Ethereum smart contracts. Our experimental results demonstrate significant improvements in scalability, development simplicity, and usability, making \textit{FlexiContracts+} a practical solution for evolving decentralized applications. Future research could extend \textit{FlexiContracts+} to enable cross-chain interoperability, strengthen its security against emerging threats, and optimize its performance for high transaction volumes, enhancing its robustness for smart contract upgradability across diverse blockchain ecosystems.

\bibliographystyle{IEEEtran}
\bibliography{IEEEabrv,main}
\end{document}